%% file: ms.tex
\PassOptionsToPackage{dvipsnames}{xcolor}
\documentclass[conference]{IEEEtran}
\IEEEoverridecommandlockouts
\usepackage{pgfplots}
\usepackage{pgfplotstable}
\usepackage{caption}
\usepackage{enumitem}
\input{macro}

\def\BibTeX{{\rm B\kern-.05em{\sc i\kern-.025em b}\kern-.08em
    T\kern-.1667em\lower.7ex\hbox{E}\kern-.125emX}}
\begin{document}

\title{LadyBug: A GitHub Bot for UI-Enhanced Bug Localization in Mobile Apps  
}

\author{
	
	\IEEEauthorblockN{Junayed Mahmud\IEEEauthorrefmark{1},
    James Chen\IEEEauthorrefmark{1},
Terry Achille\IEEEauthorrefmark{1},
Camilo Alvarez-Velez\IEEEauthorrefmark{1},
Darren Dean Bansil\IEEEauthorrefmark{1}, \\
Patrick Ijieh\IEEEauthorrefmark{1},
Samar Karanch\IEEEauthorrefmark{1},
Nadeeshan De Silva\IEEEauthorrefmark{2},
Oscar Chaparro\IEEEauthorrefmark{2},
Andrian Marcus\IEEEauthorrefmark{3},
Kevin Moran\IEEEauthorrefmark{1}},
    \IEEEauthorblockA{\IEEEauthorrefmark{1}\textit{University of Central Florida (USA)}, 
		    \IEEEauthorrefmark{2}\textit{William \& Mary (USA)},
		    \IEEEauthorrefmark{3}\textit{George Mason University (USA)}
		    \\ \href{mailto:}{junayed.mahmud@ucf.edu}, \href{mailto:}{james.chen@ucf.edu}, 
            \href{mailto:}{terry.achille@ucf.edu}, \href{mailto:}{ca840220@ucf.edu},
            \\\href{mailto:}{darrenbansil@ucf.edu},
            \href{mailto:}{patrick.ijieh@ucf.edu},
            \href{mailto:}{sa898457@ucf.edu},
            \href{mailto:}{kgdesilva@wm.edu},
           \href{mailto:}{oscarch@wm.edu},
		    \\\href{mailto:}{amarcus7@gmu.edu},
            \href{mailto:}{kpmoran@ucf.edu}
    }
}
\maketitle

\begin{abstract}
This paper introduces \textit{LadyBug}, a GitHub bot that automatically localizes bugs for Android apps by combining UI interaction information with text retrieval. LadyBug connects to an Android app's GitHub repository, and is triggered when a bug is reported in the corresponding issue tracker. Developers can then record a reproduction trace for the bug on a device or emulator and upload the trace to LadyBug via the GitHub issue tracker. This enables LadyBug to utilize both the text from the original bug description, and UI information from the reproduction trace to accurately retrieve a ranked list of files from the project that most likely contain the reported bug. 

We empirically evaluated LadyBug using an automated testing pipeline and benchmark called RedWing that contains 80 fully-localized and reproducible bug reports from 39 Android apps. Our results illustrate that LadyBug outperforms text-retrieval-based baselines and that the utilization of UI information leads to a substantial increase in localization accuracy. LadyBug is an open-source tool, available at \textit{\url{https://github.com/LadyBugML/ladybug}}.

A video showing the capabilities of Ladybug can be viewed here: \textit{\url{https://youtu.be/hI3tzbRK0Cw}}
\end{abstract}

\input{1_intro}

\input{2_tool}
\input{3_approach}
\input{4_evaluation}

\input{5_related_work}

\input{6_conclusion}

\input{7_acknowledgements}

\bibliographystyle{IEEEtran}
\bibliography{references}

\end{document}

%% file: macro.tex
\usepackage{url}
\usepackage{colortbl}
\usepackage{balance}
\usepackage{xspace}
\usepackage{graphicx}
\usepackage{verbatim}
\usepackage{bold-extra}
\usepackage{amsmath}
\usepackage{multirow}
\usepackage{listings}
\usepackage{boxedminipage}
\usepackage{soul}
\usepackage{enumitem}
\usepackage[normalem]{ulem}
\setlist{nolistsep,leftmargin=.5cm}
\usepackage{booktabs}
\usepackage{tabularx}
\usepackage{svg}
\usepackage{calc}
\usepackage{float}
\usepackage{multirow}
\usepackage[normalem]{ulem}
\useunder{\uline}{\ul}{}
\usepackage[most]{tcolorbox}
\definecolor{MidnightBlue}{HTML}{006895}
\definecolor{BoxesBlue}{HTML}{DEECFF}
\definecolor{BoxesYellow}{HTML}{FFF2CC}
\definecolor{StateGreen}{HTML}{91C788}
\definecolor{StateRed}{HTML}{FF8080}
\definecolor{ArrowGreen}{HTML}{61B15A}
\definecolor{ArrowViolet}{HTML}{BA94D1}
\usepackage{caption}
\usepackage{microtype} 
\usepackage{cite}
\usepackage{newtxtext,newtxmath}
\usepackage{algorithmic}
\usepackage{graphicx}
\usepackage{textcomp}
\usepackage[hidelinks]{hyperref}
\usepackage{ifthen}
\usepackage{wrapfig}
\usepackage{subcaption}
\usepackage{cleveref} 
\usepackage{enumitem}
\usepackage{indentfirst}
\usepackage[T1]{fontenc}

\newboolean{showcomments}
\setboolean{showcomments}{false}

\newboolean{showboxes}
\setboolean{showboxes}{true}

\usepackage{adjustbox}
\usepackage{totcount}

\ifthenelse{\boolean{showcomments}}

\makeatletter

\definecolor{mycolor}{RGB}{204,204,204}

\newcommand*\circled[1]{\tikz[baseline=(char.base)]{\small{\textbf{
      \node[shape=circle,fill=mycolor,draw=black, inner sep=0.75pt] (char) {\textcolor{black}{#1}};}}}}

\newcounter{myboxcounter}

\setboolean{showcomments}{true}

\ifthenelse{\boolean{showcomments}}
  {\newcommand{\nb}[2]{
    \fbox{\bfseries\sffamily\scriptsize#1}
    {\sf\small$\blacktriangleright$\textit{#2}$\blacktriangleleft$}
   }
   
  }
  {\newcommand{\nb}[2]{}
   
  }

\newcommand\rev[1]{{\color{black}{#1}}}



%% file: 1_intro.tex
\section{Introduction}
\label{sec:intro}

The management of bug reports is a time-consuming process for developers~\cite{murphy2013design,11025725}. One of the most difficult bug report management tasks is localizing a failure described in a bug report to the code of the corresponding project. This requires reasoning between the natural language description of a given bug and the various programming languages with which a project is written. This task is further complicated by the often inadequate or incomplete information in bug reports~\cite{Chaparro2019,11025948}.

Researchers have proposed various techniques to automate the bug localization process. However, a large body of this prior work tends to formulate this problem as a text retrieval~(TR)~\cite{davies2012using} problem, wherein a bug report is treated as a query and different granularities of source code (files, classes, methods, etc.) are ranked and presented to the programmer based on their likelihood of containing the described bug. 

The primary assumption of TR-based bug localization is that bug reports and the corresponding source code will share terms; however, there exists a semantic gap between the information present in bug reports and the information available in source code~\cite{Moran2015}. Previous research proposed different techniques to mitigate this semantic gap in existing TR-based bug localization techniques. For instance, researchers have introduced techniques to preprocess the text in bug reports or source code,  
 while other techniques focus on query reformulation utilizing information from various resources~\cite{Rahman2018}. Other lines of research focused on boosting the ranking of buggy code elements using execution information (e.g., stack traces~\cite{youm2017improved}), code dependencies (e.g., static code analysis~\cite{dit2013feature}), or app version history information (e.g., mining from GitHub repositories~\cite{youm2017improved}).

In this paper, we propose a novel tool for bug localization of Android applications called LadyBug, which ranks buggy source code files using the bug descriptions from reports \textit{and} a new, largely underexplored source -- information from an app's graphical user interface (GUI). The key intuition that underlies LadyBug is that manually or automatically collected bug reproduction traces which capture GUI-level interactions for recreating a bug can be utilized to help \textit{refine} the localization process (i.e., filtering out files unlikely to contain the bug while boosting the rankings of files that are more likely to contain the bug.) As such LadyBug uses GUI interaction data in conjunction with a neural text embedding approach (i.e., UniXCoder~\cite{guo2022unixcoder}), to generate a ranked list of files based on the likelihood of containing bugs. 

We constructed LadyBug as a GitHub bot that can be easily added to GitHub repositories of Android applications and used as an automated assistant for bug localization. LadyBug's GitHub bot architecture consists of three key components: (i) a repository connector, built on top of the ProBot~\cite{probot} framework, which allows for Ladybug to post/receive information to/from GitHub issues, (ii) a backend that indexes a given repository and implements the text retrieval engine and (iii) \textit{MetaMorph}, a desktop application which allows a developer to easily record reproduction traces of a described bug. LadyBug can also be used in conjunction with recent automated reproduction tools~\cite{wang:issta25} which automates the localization task end-to-end. 

\begin{figure*}[t]
  \centering
  \vspace{-1em}
  \includegraphics[width=1\textwidth]{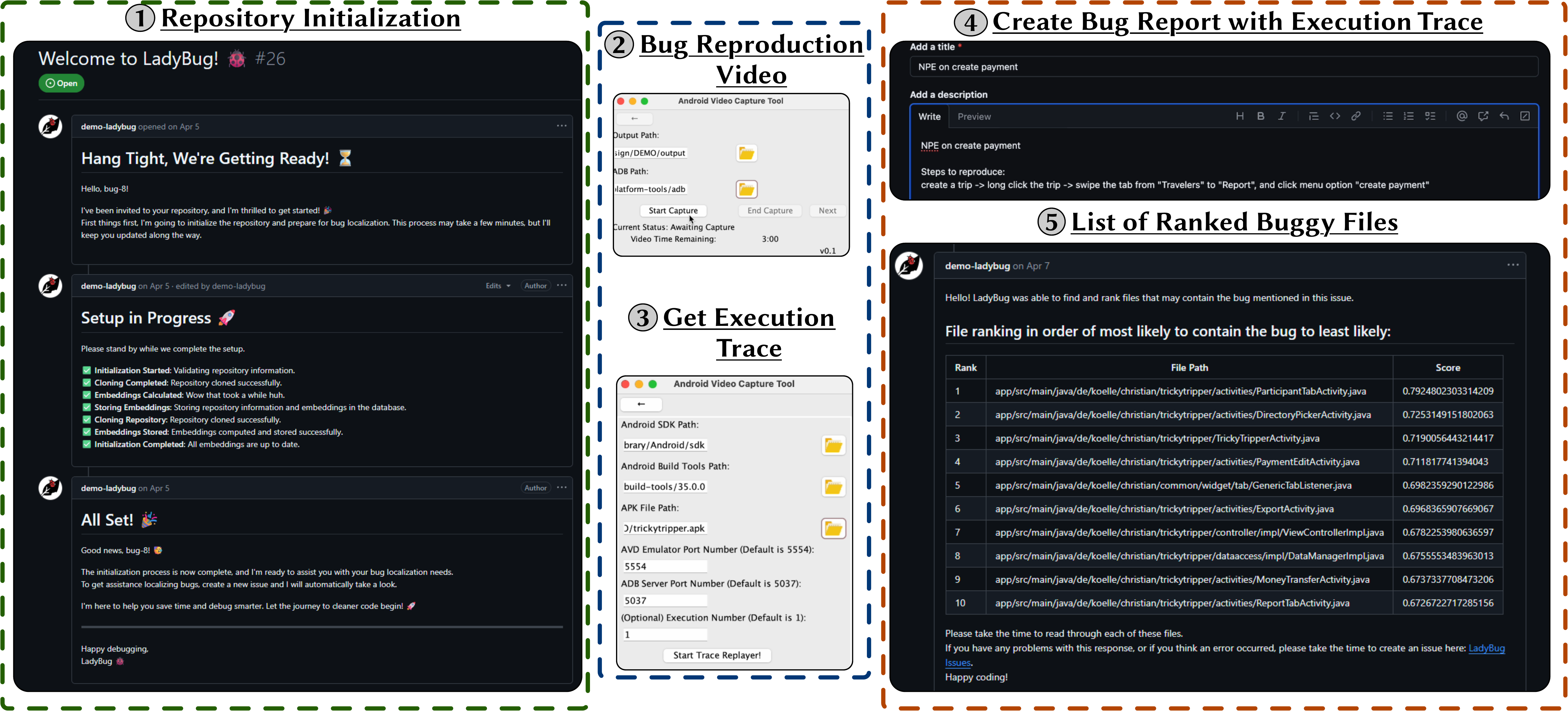}
  \caption{The Typical User Workflow of the LadyBug GitHub Bot}
  \vspace{-1em}
  \label{fig:user-diagram}
\end{figure*}

We empirically evaluated LadyBug using a benchmark called RedWing that contains 80 bug reports from 39 Android applications compared to a text retrieval baseline. Our results illustrate that (i) using GUI information improves the accuracy of file-level bug localization for Android apps, and (ii) LadyBug is able to outperform a text-only baseline. LadyBug is an open-source tool~\cite{ladybug-repo}, and is compatible with any Android app. Further details about LadyBug's algorithms and an expanded evaluation can be found in its original research paper~\cite{Mahmud:ICSE2024}.

%% file: 2_tool.tex
\section{LadyBug: A GUI Bug Localization Tool for Android Applications}
\label{sec:tool}

LadyBug is a GitHub bot that automates bug localization by analyzing Java source code and Native Android GUI interactions, reporting results directly within GitHub issue threads. 
LadyBug workflow is structured into three primary phases, illustrated in Figure~\ref{fig:user-diagram}. The first is the \textit{Repository Initialization} phase (highlighted in {\color{ForestGreen}green}), where the Ladybug bot is installed on the target GitHub repository and granted the necessary permissions to monitor repository activities. Second, during the \textit{Bug Reproduction} phase (highlighted in {\color{NavyBlue}blue}), the MetaMorph desktop application assists a developer with recording and replaying a reproduction trace of a given bug, generating an execution trace capturing detailed GUI interactions. Lastly, the \textit{Issue Reporting} phase (highlighted in {\color{Bittersweet}orange}) involves \rev{developers} submitting new GitHub issues to trigger the bug localization process, with an option to include the previously recorded execution trace to aid localization.

\subsection{Repository Initialization}
The first thing that a \rev{developer} will need to do is add the Ladybug bot to their GitHub repository through the GitHub front-end (see Figure~\ref{fig:user-diagram}-\circled{1}). After adding LadyBug to a given repository, LadyBug will create a new issue giving the \rev{developer} a welcome message and incremental updates on the initialization process. When all these steps are completed, a confirmation message will be posted to the issue thread, indicating that LadyBug is ready for further use.

\subsection{Bug Reproduction}
When the MetaMorph application is opened (Figure~\ref{fig:user-diagram}-\circled{2}), the \rev{developer} will be tasked with providing system paths to the Android emulator's files, as well as system paths to the Android application. After the \rev{developer} has completed providing the information requested, they are asked to open the application inside of their emulator and press "Start Capture". Then, the \rev{developer} will retrace the steps described in the original bug report, performing actions until the bug occurs in the app on the emulated device -- and then hit "Stop Capture". Afterwards, the \rev{developer} will supply the necessary information MetaMorph needs to retrace the steps the \rev{developer} took in the bug report, and click on "Start Trace Replayer" (Figure~\ref{fig:user-diagram}-\circled{3}). After this is done, MetaMorph writes this information into an execution.json file, which contains the necessary GUI information that LadyBug needs to perform GUI-enhanced bug localization.
\looseness=-1

\subsection{Issue Reporting}
With all the necessary GUI information, the \rev{developer} can create a new GitHub issue with the trace and bug report (Figure~\ref{fig:user-diagram}-\circled{4}). LadyBug will respond to the issue with status messages updating the user on the bug localization process. After localizing the bug, the LadyBug bot will reply with the ranked list of potential buggy files (Figure~\ref{fig:user-diagram}-\circled{5}).

%% file: 3_approach.tex
\section{LadyBug's Architecture \& Implementation}
\label{sec:architecture}
\begin{figure*}[t]
    \centering
    \vspace{-2.5em}
    \includegraphics[width=.88\textwidth]{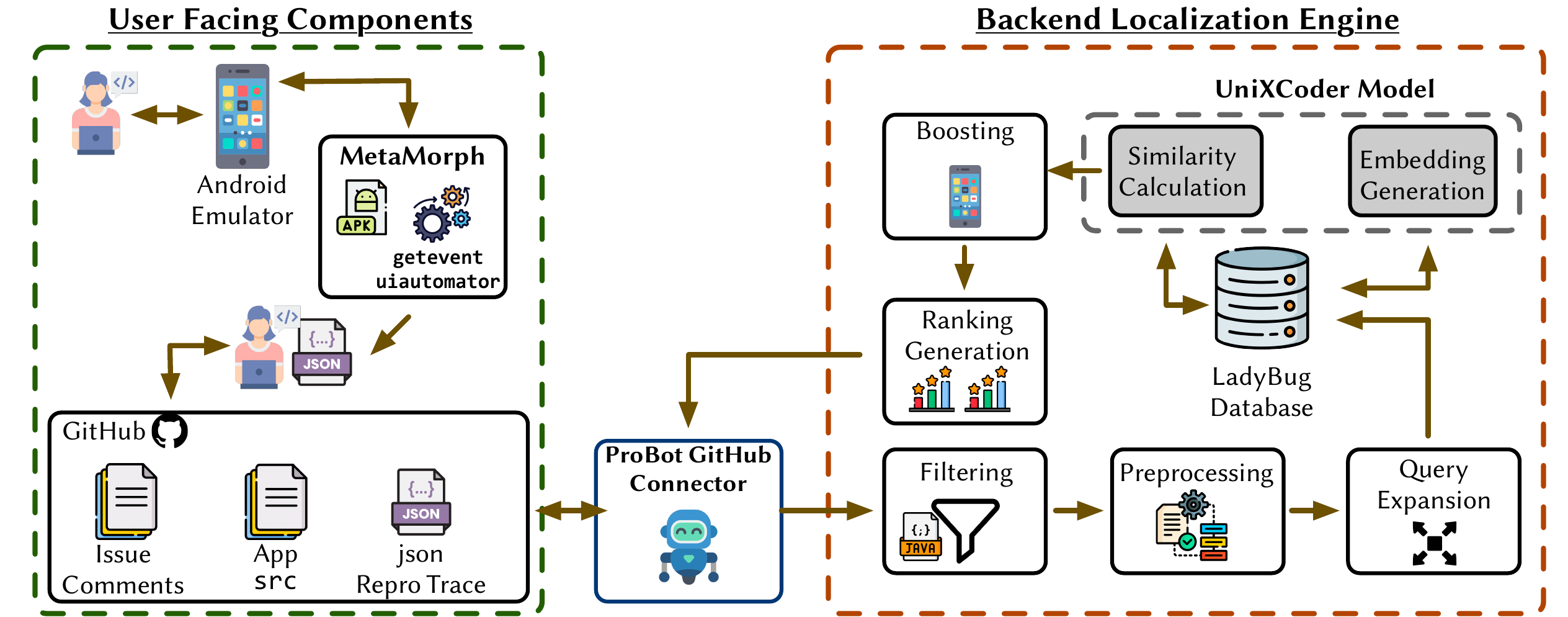}
    \caption{Ladybug's Architecture}
    \vspace{-1.5em}
    \label{fig:arch}
\end{figure*}

Ladybug is comprised of three main components shown in Figure~\ref{fig:arch}: (i) the MetaMorph desktop application for Mac/Windows (ii) the GitHub connector that facilitates communication between the Localization engine and GitHub, and (iii) the backend server that implements the bug localization engine.

\subsection{The MetaMorph Bug Reproduction Tool}
While LadyBug can be used in conjunction with automated tools that reproduce bugs on an Android device given a bug report~\cite{wang:issta25}, these tools are not perfect, and some developers may prefer to reproduce bugs manually. As such, LadyBug is packaged with MetaMorph, a Mac and Windows Desktop application that facilitates the recording of a reproduction trace from an Android device or emulator. MetaMorph has two main components, a \textit{recorder} and a \textit{replayer}. The \textit{recorder} makes use of the \texttt{\small\textbf{getevent}} utility, included as a part of the Android kernel, that is capable of capturing fine-grained touch information with millisecond-level accuracy and the Android  \texttt{\small\textbf{screencap}} utility to record both a trace and video of a given bug. Then the \textit{replayer} takes as input the \texttt{\small\textbf{getevent}} trace, and automatically replays it on the device, this time capturing fine-grained GUI information of each widget that the user interacted with to reproduce the bug using the Android \texttt{\small\textbf{uiautomator}} framework. The \textit{replayer} exports a json file that contains all of the GUI interaction information needed by LadyBug.

\vspace{-0.5em}
\subsection{ProBot GitHub Repository Connector}
\vspace{-0.5em}
LadyBug's Repository connector is built on the Probot framework, which offers a high-level abstraction for developing GitHub Apps using Node.js and the Octokit library. Probot listens for webhook events—such as repository installations and issue creations—and routes them through an internal handler, allowing the app to respond programmatically via GitHub’s REST API. LadyBug uses this infrastructure to authenticate as a GitHub App, monitor repository events, and interface directly with GitHub issues. The backend logic, written in JavaScript, handles routing, repository tracking, and progress updates.

When LadyBug is first installed on a repository, it records the repository identifier ("owner/repo") and the latest commit SHA, then creates a welcome issue to initiate setup. This SHA and the repository’s source code are stored in the LadyBug database in the backend server, and embeddings are computed using the UniXCoder model in preparation for retrieval of source code files pending an incoming issue. We chose to use the UniXCoder model as it was one of the better performing models reported in the original technical paper~\cite{Mahmud:ICSE2024}. However, LadyBug is extensible to any text-embedding format, including from Large Language Models (LLMs).
When a user opens a new GitHub issue to report a bug, LadyBug validates the request and posts an initial comment to confirm that the report has been received. If the report includes a reproduction trace, it is parsed and sent to the backend along with the issue content and repository metadata. Before proceeding, LadyBug verifies whether the commit SHA in the report matches the stored version. If it does not, the initialization phase is re-triggered to ensure the GitHub source code files and computed embeddings are current. The updated issue data is then compared against the stored embeddings to generate a ranked list of likely buggy files. The final ranked output is posted as a comment to the GitHub issue.
\looseness=-1

\vspace{-0.5em}
\subsection{Localization Engine}
\vspace{-0.5em}
LadyBug's localization engine is a Flask server fully written in Python that handles bug localization. It listens for two requests sent from Probot: initialization and report. Probot sends an `initialization' request when LadyBug is added to a repository and a `report' request whenever a GitHub Issue is created on the target repository. 

Upon an `initialization' request, LadyBug, using its Probot credentials, clones the repository’s default branch. LadyBug filters the repository to retain only Java source files—any files or directories not matching the .java extension are discarded. Each remaining Java files go through preprocessing for uniform representation and their respective embeddings are stored in LadyBug's embedding database. LadyBug performs the following preprocessing steps: (i) Sanitization: Strips comments, imports, and special characters not relevant to semantics. (ii) Tokenization: Breaks the source text into syntactic tokens (identifiers, keywords, literals). (iii) Normalization: Removes language-specific stop tokens (e.g., punctuation, boilerplate keywords) and applies lemmatization to identifiers and literals. (iv) Segmentation: Splits long files into semantically coherent segments if they exceed 500 tokens.

When the localization engine receives a `report' request, LadyBug will begin by extracting relevant GUI information from the attached execution.json file. Using regex, resource-ids of individual GUI components are stored and classified as \textbf{Screen Component terms}. Activity/Fragment names are stored and classified as \textbf{GUI Screen terms}.  Screen Component terms are appended to the end of the bug report and then used for \textbf{Query Reformulation}. After this, a corpus of source code file embeddings is built using a brute-force approach: If the source code file contained any Screen Component Term, \textbf{Filtering} is performed to remove files not related to the relevant screens. 
Lastly, \textbf{Boosting} is performed by boosting source code files that contain GUI Screen terms to the top of the ranked list, preserving relative order. This configuration of Filtering, Boosting, and Query Reformulation were derived from the best performing configuration in LadyBug's expanded evaluation in it's technical paper~\cite{Mahmud:ICSE2024}.

%% file: 4_evaluation.tex
\section{Evaluation}
\label{sec:evaluation}
We conducted an empirical study to understand the impact of using GUI interaction information on text-retrieval-based bug localization. We formulated the following research question:
\begin{itemize}
    \item[]{\textbf{RQ:} \textit{Does the use of GUI Interaction information improve text-retrieval-based bug localization for Android apps?}}
\end{itemize}
\vspace{-1em}
\subsection{Data}
To answer RQ, we developed \textit{Redwing}, an automated testing suite designed specifically to assess the effectiveness of LadyBug's localization methods, and to ensure that we were able to replicate the results of the larger empirical study performed in LadyBug's technical paper~\cite{Mahmud:ICSE2024}. We used the same set of 80 reproducible bug reports from 39 different open-source Android applications used in LadyBug's technical paper. Each bug report includes the description of the bug, metadata for the bug, and the recorded steps/scenarios for reproducing the bug. For a full description of the data collection methodology, please see the technical paper~\cite{Mahmud:ICSE2024}.

\subsection{Methodology}
Redwing leverages the core bug localization pipeline, adapting it to systematically process a predefined dataset using known buggy files as ground truth. By comparing the pipeline's predicted outcomes against these established ground truths, Redwing calculates several widely recognized evaluation metrics, including (i) Hits@K: indicates the frequency with which the correct buggy file appears within the top K results. A higher value suggests that the system consistently narrows down bugs to a manageable subset of files. (ii) Mean Reciprocal Rank (MRR): Measures how quickly the first buggy file is identified. A higher score (closer to 1) indicates better performance, meaning fewer irrelevant files to review before locating the correct one. (iii) Mean Average Precision (MAP): Evaluates the overall ranking accuracy of all buggy files within the results. Higher MAP scores demonstrate the system's precision in consistently placing the correct files near the top of the list. (iv) Effectiveness (E): Represents the rank position of the first identified buggy file, where lower ranks are preferable. This metric directly reflects the practical efficiency, indicating how many files developers must review before pinpointing the bug.

The Redwing testing suite integrates essential components from both the initialization and reporting phases of the main localization pipeline, resulting in a cohesive and streamlined evaluation process. This consolidation eliminates performance bottlenecks typically associated with external dependencies, such as API calls, data streaming, and manual user inputs, thus ensuring efficient and reproducible testing.
As a command-line interface (CLI) tool, Redwing provides users flexibility and ease of use, allowing users to specify various parameters, including the dataset file path, localization type (with GUI data, without GUI data, or a comparative mode to calculate the relative improvement), and the number of asynchronous test iterations to execute. This modular and user-friendly design supports comprehensive and customizable experimentation. 

\subsection{Results}

\begin{table}[t]
    \setlength\tabcolsep{3.5pt}
    \small
  \centering
  \caption{Metrics for GUI-augmented vs. text-only (No GUI) data.}
  \label{tab:metrics}
  \resizebox{0.98\columnwidth}{!}{%
  \begin{tabular}{c|c|c|c|c|c|c}
    \hline
    \textbf{Augmentation} & \textbf{H@1} & \textbf{H@5} & \textbf{H@10} & \textbf{MAP} & \textbf{MRR} & \textbf{Effectiveness}\\
    \hline
    UniXCoder-GUI & 0.30& 0.74& 0.81& 0.39 & 0.44& 13.81\\
    UniXCoder-No GUI & 0.20 & 0.61 & 0.68 & 0.31 & 0.36 & 16.41\\
    \hline
  \end{tabular}
  }
  \vspace{-2em}
\end{table}

In Table~\ref{tab:metrics} presented above, it is evident that incorporating text-based retrieval techniques as an augmentation to the GUI-based data consistently enhances retrieval performance across all evaluated values of k in the Hits@k metric (H@K). Notably, our primary focus was on the Hits@10 metric, as \rev{this represents a manageable number of files for a developer to examine~\cite{Rahman2018, Wen2016}}. By integrating text-based retrieval with GUI data, we observed a substantial improvement in the Hits@10 score, achieving a relative increase of 16.2\%. These results suggest that the complementary nature of textual information in enriching the semantic understanding of GUI representations facilitates more accurate and relevant retrieval outcomes. Additionally, these results replicate the results for the same configuration of LadyBug in its corresponding technical paper, ensuring that our implementation of the localization in our LadyBug GitHub bot, matches that of the original algorithms developed in the technical paper~\cite{Mahmud:ICSE2024}.

In addition, augmenting the GUI data with additional retrieval signals has led to noticeable improvements in both the MAP and MRR metrics. Specifically, the rise in MAP suggests a greater overall precision across multiple queries. At the same time, the higher MRR values reflect a more consistent placement of the correct file near the top of the ranked list. In terms of effectiveness, we also see that with GUI data, the mean efficacy decreases. This suggests that using GUI data, we can more accurately rank files based on their relevance to the bug. These findings demonstrate that using GUI data contributes to a more effective retrieval process by reducing the prominence of irrelevant files at the top of the list.

%% file: 5_related_work.tex
\section{Related Work}
\label{sec:related_work}
Text-Retrieval-based Bug Localization (TRBL) techniques use text similarity between code and bug reports to identify potentially buggy code artifacts. A relevance score determines how likely code is to contain bugs.
TRBL techniques fall into three main categories: classical, DL, and combined.
\looseness=-1

\textit{Classical IR techniques} employ algorithms including the Unigram Model (UM), Cluster Based Document Model (CBDM), Vector Space Model (VSM), Latent Semantic Indexing (LSI), and Latent Dirichlet Allocation (LDA).

\textit{DL approaches} utilize neural network architectures including Convolutional Neural Networks (CNNs), Recurrent Neural Networks (RNNs), Transformers, and hybrid models that combine these architectures.

Other approaches combine IR and DL techniques \cite{Lam2017}. 
\looseness=-1

To improve bug localization accuracy by bridging the gap between bug reports and source code, researchers incorporate additional information sources including similar bug reports \cite{rath2018analyzing}, code structure \cite{takahashi2018preliminary}, version history \cite{zhang2019commit}, stack traces \cite{Wen2016}, and part-of-speech data \cite{zhou2017augmenting}.

LadyBug is novel compared to such prior work as we incorporate GUI interaction data from Android apps to bridge the lexical gap between bug reports and source code, thereby improving buggy file retrieval.

Prior work has employed various augmentation strategies to improve the ranking of buggy code artifacts in TRBL, including boosting, filtering, and query (re)formulation. Boosting increases the relevance scores of artifacts to rank buggy ones higher \cite{lou2021boosting}, while filtering removes irrelevant artifacts from the search space or initial results \cite{liu2007feature}. Query (re)formulation modifies queries to better capture relevant information through expansion (adding terms) \cite{kim2021novel}, replacement (substituting the query) \cite{guo2017tackling}, and reduction (removing unhelpful terms) \cite{ florez2021combining}. Building on this work, we assess the impact of incorporating GUI interaction data through boosting, filtering, query expansion, and replacement, and plan to explore its integration with query reduction in future work.

LadyBug employs boosting, filtering, and query expansion as augmentation techniques with a new source of data, namely, GUI interaction data. Our adaption of these techniques to GUI interaction data in Android apps is novel.

%% file: 6_conclusion.tex
\section{Final Remarks \& Future Work}
\label{sec:conclusion}
Ladybug is a tool that helps Android developers accurately localize bugs in their source code by combining text retrieval with boosting and filtering strategies utilizing GUI interaction data. By leveraging language model to encode both source files and user-reported bug descriptions, and by extracting resource-id and view-hierarchy terms from replayed interaction traces, Ladybug generates a ranked list of candidate files that more likely to be buggy than traditional techniques. In the future, we plan on expanding the amount of embedding models used to create source code and bug report embeddings. This would ensure that both code and natural-language inputs benefit from the latest advances in representation learning. We also intend to enrich Ladybug’s user-facing output: (e.g., code fix suggestions, bug explanations). \rev{We will also enhance the design of LadyBug to accept the MetaMorph results at any time while an issue remains open.}
\looseness=-1

%% file: 7_acknowledgements.tex
\section*{Acknowledgement}

This research has been supported in part by the NSF CCF-2441355, CCF-2007246, and CCF-1955853. The authors also acknowledge funding support from Apple. Any opinions, findings, and conclusions expressed herein are the authors’ and do not necessarily reflect those of the sponsors.